\documentclass[galaxies,article,accept,moreauthors,pdftex]{mdpi}
\firstpage{1} 
\pubvolume{1}
\issuenum{1}
\articlenumber{0}
\pubyear{2023}
\copyrightyear{2023}
\externaleditor{{Academic Editor: Oleg Malkov} 
}
\datereceived{14 March 2023} 
\daterevised{5 June 2023} % Only for the journal Acoustics
\dateaccepted{12 June 2023} 
\datepublished{} 
\hreflink{https://doi.org/} % If needed use \linebreak
\usepackage[defaultcolor=red]{changes}

\makeatletter
\@namedef{Changes@AuthorColor}{red}
\colorlet{Changes@Color}{red}
\makeatother

\newcommand{\aap}{Astron. Astrophys.}
\newcommand{\aaps}{Astron. Astrophys. Suppl. Ser.}
\newcommand{\aapr}{Astron. Astrophys. Rev.}

\newcommand{\aj}{Astron. J.}
\newcommand{\apj}{Astrophys. J.}
\newcommand{\apjl}{Astrophys. J.}
\newcommand{\apjs}{Astrophys. J. Suppl.}
\newcommand{\apss}{Astrophys. Space Sci.}
\newcommand{\jaavso}{J. Am. Assoc. Variable Star Observers}
\newcommand{\mnras}{Mon. Not. R. Astron. Soc.}

\newcommand{\nat}{Nature}
\newcommand{\pasa}{Publ. Astron. Soc. Aust.}

\newcommand\farcs{\mbox{$.\!\!^{\prime\prime}$}}% 
\let\ga=\gtrsim%

\Title{Dense Molecular Environments of B[e] Supergiants and \linebreak Yellow Hypergiants}
\TitleCitation{Dense Molecular Environments of B[e] Supergiants and  Yellow Hypergiants}

\Author{{Michaela Kraus} 
 $^{1,}$*\orcidA{}, Michalis Kourniotis $^{1}$\orcidB{}, {Mar\'{i}a Laura Arias} 
 $^{2,3}$\orcidC{}, {Andrea F. Torres} $^{2,3}$\orcidD{} and {Dieter H. Nickeler} $^{1}$\orcidE{}}

\AuthorNames{Michaela Kraus, Michalis Kourniotis, Mar\'{i}a Laura Arias, Andrea F. Torres  and Dieter H. Nickeler}

\AuthorCitation{Kraus, M.; Kourniotis, M.; Arias, M.L.; Torres, A.F.; Nickeler, D.H.}

\address{%
$^{1}$ \quad Astronomical Institute, Czech Academy of Sciences, 
 Fri\v{c}ova 298, 251\,65 Ond\v{r}ejov, Czech Republic; 
\\ 
$^{2}$ \quad Departamento de Espectroscop\'ia, Facultad de Ciencias Astron\'omicas y Geof\'isicas, Universidad Nacional de La Plata, Paseo del Bosque S/N, {La Plata} 
  B1900FWA, Argentina; \\
 $^{3}$ \quad Instituto de Astrof\'isica de La Plata (CCT La Plata-CONICET, UNLP)
Paseo del Bosque S/N, \linebreak {La Plata} B1900FWA, Argentina\\
 }

\corres{Correspondence: michaela.kraus@asu.cas.cz}

\abstract{Massive stars expel large amounts of mass during their late evolutionary phases. We aim to unveil 
the physical conditions within the warm molecular environments of B[e] supergiants (B[e]SGs) and 
yellow hypergiants (YHGs), which are known to be embedded in circumstellar shells and disks. We present 
$K$-band spectra of two B[e]SGs from the Large Magellanic Cloud and four Galactic YHGs. The CO band 
emission detected  from the B[e]SGs LHA~120-S~12 and LHA~120-S~134 suggests that these stars are 
surrounded by stable rotating molecular rings. The spectra of the YHGs display a rather diverse 
appearance. The objects 6~Cas and V509~Cas lack any molecular features. The star [FMR2006]~15 displays 
blue-shifted CO bands in emission, which might be explained by a possible close to pole-on oriented 
bipolar outflow. In contrast, HD~179821 shows blue-shifted CO bands in absorption. While the star 
itself is too hot to form molecules in its outer atmosphere, we propose that it might have experienced 
a recent outburst. We speculate that we currently can only see the approaching part of the expelled 
matter because the star itself might still block the receding parts of a (possibly) expanding gas 
shell. }


\begin{document}

%%%%%%%%%%%%%%%%%%%%%%%%%%%%%%%%%%%%%%%%%%

\section{Introduction}

The evolution of massive stars ($M_{\rm ini} \ga 8$\,M$_{\odot}$) bears many uncertainties, 
which render it difficult to trace such objects from the cradle up to their spectacular 
explosion as supernova. One major hindrance is the poorly constrained mass loss due to stellar 
winds that the stars experience along the course of their evolution. Furthermore, the post-main 
sequence evolution of massive stars encounters phases in which the stars  {lose} a significant 
amount of mass due to episodically enhanced mass loss or occasional mass eruptions, both 
of poorly understood origin. The ejected mass can accumulate around the star in rings, 
shells, or bipolar lobes, as seen in some B- or B[e]-type supergiants (e.g., Sher 25 
\cite{1997ApJ...475L..45B, 2008MNRAS.388.1127H}, MWC\,137 \cite{2008A&A...477..193M, 
2017AJ....154..186K}, SBW1 \cite{2013MNRAS.429.1324S}), yellow hypergiants (IRC+10 420 
\cite{2013A&A...551A..69O} and Hen 3-1379 \cite{2011A&A...534L..10L}), many luminous blue 
variables \cite{2011IAUS..272..372W, 2022Galax..10...41L}, and Wolf-Rayet stars 
\cite{1981ApJ...249..195C, 1994ApJS...93..229M, 1994ApJS...95..151M, 2020A&A...635A.201M, 
2021MNRAS.501.5350S}. 

Two groups of evolved massive stars are particularly interesting. These are the B[e] 
supergiants (B[e]SGs) and the yellow hypergiants (YHGs). Both types of objects have dense and 
warm circumstellar environments, and representatives of both classes of objects show (at least 
occasionally) emission from hot molecular gas.

\subsection{B[e] Supergiants}

The group of B[e]SGs consists of luminous ($\log (L_{*}/L_{\odot}) \geq 4.0$) post-main 
sequence B-type emission line stars. Besides large-scale ejecta (with sizes of several pc) 
detected in some B[e]SGs \cite{2008A&A...477..193M, 2022Galax..10...41L}, all objects have 
intense winds and are surrounded by massive disks on small scales (up to $\sim$100 AU) 
\cite{1986A&A...163..119Z, 2007A&A...464...81D, 2018MNRAS.480..320M}, giving rise to the 
specific emission features characterizing stars with the B[e] phenomenon 
\cite{1998A&A...340..117L}. These disks give shelter to a diversity of molecular and dust 
species, and the near-infrared (NIR) is an ideal wavelength regime to detect molecular 
emission features that, when resolved with high spectral resolution, provide insight into the 
physical properties of the disks and reveal the disk dynamics. 

The most commonly observed molecule is CO. The emission from its first-overtone bands arises in the 
$K$-band around $2.3\,\upmu$m and has been detected in about 50\% of the \mbox{B[e]SGs 
\cite{2019Galax...7...83K}.} Besides the main isotope $^{12}$CO, emission from $^{13}$CO 
is seen in considerable amounts \cite{2010MNRAS.408L...6L, 2013A&A...558A..17O}, confirming 
that the matter from which the disks have formed contains processed material that must have 
been released from the stellar surface \cite{2009A&A...494..253K}. 

Emission from the first-overtone bands of SiO, arising in the $L$-band around $4\,\upmu$m, has 
been reported for some Galactic B[e]SGs \cite{2015ApJ...800L..20K}. SiO has a lower binding 
energy than CO. It thus forms at distances farther away from the star than CO. The individual 
ro-vibrational lines in both CO and SiO are kinematically broadened with a double-peaked 
profile, and (quasi-)Keplerian rotation of the molecular gas around the central object 
has been suggested as the most likely explanation to interpret the spectral appearance 
\cite{2015ApJ...800L..20K, 2016A&A...593A.112K, 2013A&A...549A..28K, 2015AJ....149...13M, 
2018MNRAS.480..320M, 2018A&A...612A.113T, 2012A&A...548A..72C}.

Observations of B[e]SGs in the NIR are sparse. But persistent CO band emission over years and 
decades has been detected in numerous objects \cite{1988ApJ...334..639M, 1988ApJ...324.1071M, 
2013A&A...558A..17O} and has been used as one of the criteria to identify and classify stars 
as B[e]SGs in Local Group \mbox{galaxies \cite{2014ApJ...780L..10K, 2015MNRAS.447.2459S}.} However, 
in a few cases, considerable variability in these emission features has been reported as well. 
The most striking object is certainly the B[e]SG star LHA 115-S 65 
in the Small Magellanic Cloud (SMC), for which a sudden appearance of CO band emission has been 
recorded \cite{2012MNRAS.426L..56O}. The disk around this object is seen edge-on, and in 
addition to its rotation around the central object, it also drifts outwards, very slowly and
with a velocity decreasing with distance from the star and reaching about zero 
\cite{2010A&A...517A..30K}. This slowdown might have resulted in a build-up of density in 
regions favorable for molecule condensation and for excitation of the CO bands.

Furthermore, LHA 115-S 18 in the SMC showed no CO band emission back in 1987/1989 
\cite{1989A&A...223..237M}, whereas follow-up observations in November 1995 taken with a more  
than three times higher resolution displayed intense CO bands \cite{1996ApJ...470..597M}, 
which were also seen in the observations acquired in October 2009 \cite{2013A&A...558A..17O}.

In the spectrum of the Galactic B[e]SG MWC\,349, intense CO band emission appeared in the early 1980s
\cite{1986ApJ...311..909H}. It was still observable in 2013, but by then the CO gas had clearly cooled 
and the emission intensity had significantly decreased, which has been interpreted as due to 
expansion and dilution of the circumstellar disk \cite{2020MNRAS.493.4308K}. Two more objects 
in the Large Magellanic Cloud (LMC) displayed indications of CO band variability, most likely 
related to inhomogeneities within the distribution of the molecular gas around the central 
star. These are LHA 120-S 73 \cite{2016A&A...593A.112K} and LHA 120-S 35 
\cite{2018A&A...612A.113T}. 

In the optical range, indications for emission from TiO molecular bands have been found in
six B[e]SGs \cite{1989A&A...220..206Z, 2012MNRAS.427L..80T, 2016A&A...593A.112K, 
2018A&A...612A.113T}. All six objects reside in the Magellanic Clouds, and five of them also have 
CO band emission\endnote{The sixth object is LHA 120-S 111. To our knowledge, it has 
been observed in the $K$-band only once, in January 1987 \cite{1988ApJ...334..639M}. At that 
time, no CO band emission was detected.}. No Galactic B[e]SG has been reported to date to display 
TiO band emission \cite{2019Galax...7...83K}.

\subsection{Yellow Hypergiants}

With temperatures in the range $T_{\rm eff} \simeq 4000 - 8000$\,K and luminosities  
$\log (L/L_{\odot})$ spreading from $5.2$ to $5.8$, the YHGs populate a rather narrow domain 
in the Hertzprung-Russel (HR) diagram. The stars are in their post-red supergiant (post-RSG) 
evolutionary phase \cite{2009ASPC..412...17O}, and their luminosities place them on 
evolutionary tracks of stars with initial masses in the range $M_{\rm ini} \simeq
20-40$\,M$_{\odot}$. Evolutionary calculations of (rotating) stars in this mass range have shown 
that these objects can indeed evolve back to the blue, hot side of the HR diagram 
\cite{2012A&A...537A.146E}, whereas stars with lower initial mass just reach the RSG 
stage before they explode as SNe of type II-P. Support for this theoretical scenario is 
provided by the lack of high-mass ($M_{\rm ini} \geq 18$\,M$_{\odot}$, i.e., with luminosities 
$\log L/L_{\odot} > 5.1$) RSG progenitors for this type of supernovae 
\cite{2009MNRAS.395.1409S, 2015PASA...32...16S}.

As post-RSGs, the YHGs might be expected to be embedded in envelopes, remnants of the previous 
mass-losing activities during the RSG stage. However, surprisingly, so far, only about half of the YHGs 
have been reported to have a dusty and/or cold molecular envelope. These are the Galactic 
objects IRC~+10420 \cite{1996MNRAS.280.1062O, 2001A&A...368L..34C, 2010AJ....140..339T}, 
HR~5171A \cite{1990A&A...230..339D}, HD~179821 \cite{1999ApJ...525L.113J} and Hen~3-1379 
\cite{2011A&A...534L..10L, 2020A&A...635A.183K}, three YHGs in the LMC (HD~269953, HD~269723, 
HD~268757, \cite{2022MNRAS.511.4360K}), as well as Var A \cite{2006AJ....131.2105H} and three 
more YHG candidates in the galaxy M33 \cite{2017A&A...601A..76K}.

A typical classification characteristic of YHGs is the occurrence of outbursts that can be clearly 
discriminated from the more regular (cyclic) brightness variability due to stellar pulsations. During 
such an outburst event, the star inflates, its brightness drastically decreases, and the object seems 
to undergo a red loop evolution in the HR diagram. Molecules such as TiO and CO can form 
in the cool, outer atmospheric layers, leading to intense absorption structures in the optical 
and NIR, respectively, and the object's entire spectral appearance resembles that of a much 
later spectral type. The outbursts are most likely connected with enhanced mass loss or mass 
eruptions from the star, which might be connected to non-linear instabilities such as finite-time 
singularities or blow-ups typically occurring in fluid dynamics \cite{2008ASTRA...4....7N} or to 
strange-mode instabilities \cite{1994MNRAS.271...66G, 1999MNRAS.303..116G} as recent computations  
propose \cite{2022IAUS..366...51K}. The duration of the outbursts can range from months to decades 
before the star appears back at its real location in the HR diagram.

The bona-fide YHG, $\rho$~Cas, experienced four documented outbursts during the past 80 years 
with variable duration (from weeks up to three years) and amplitude (0.29 to 1.69\,mag), 
connected with significant changes in its spectral appearance \cite{2003ApJ...583..923L, 
2018ARep...62..623K, 2019MNRAS.483.3792K, 2022JAVSO..50...49M}, whereas Var A in M33 presumably 
underwent an eruption around 1950 that lasted $\sim$45\,years \cite{2006AJ....131.2105H}. The 
object V509~Cas experienced mass-loss events in the seventies, during which the star's 
apparent temperature decreased significantly. Since then, the star has displayed a steady increase 
in its effective temperature from $\sim$5000\,K in 1973 to $\sim$8000\,K in 2001 
\cite{2012A&A...546A.105N}, and since then it has stabilized at that temperature 
\cite{2017ASPC..508..239A}. Furthermore, IRC~+10420 changed its spectral type from F8 to mid- to early 
A, connected with an increase in temperature over a period of about 30 years with an average 
rate of $\sim$120\,K per year \cite{1998A&AS..129..541O, 2002ARep...46..139K}. The light 
curves of other YHG candidates also display outburst activity in connection with variable 
mass loss \cite{2017A&A...601A..76K, 2022MNRAS.511.4360K}. However, in many cases, the mass-loss 
episodes appear to be short, so that the released material expands and dilutes without 
creating detectable large-scale circumstellar envelopes \cite{2006AJ....131..603S}. 
 
Nevertheless, many YHGs are surrounded (or were for some period in the past) by hot molecular 
gas traced by first-overtone CO band emission, suggesting that the objects are embedded in a 
dense and hot environment. Whether the molecular gas is arranged in a ring revolving around the 
object, as in the case of the B[e]SGs, is currently not known. However, the CO band spectral 
features in YHGs seem to be much more variable than in their hotter B[e]SG counterparts, 
especially because they often appear superimposed on photospheric CO band absorption that 
forms during the expansion and cooling periods of the long-term pulsation cycles, especially of the 
cooler YHGs, or during outburst events. One such candidate with cyclic CO band variability is 
$\rho$~Cas. During its pulsation cycles, CO band emission appears when the star is hottest (maximum 
brightness) and most compact, whereas CO bands are seen in absorption when the star is coolest 
(minimum brightness) and most inflated \cite{2006ApJ...651.1130G}. It has been speculated that the 
appearance of CO bands in emission might be related to propagating pulsation-driven shock waves in the 
outer atmosphere of the star \cite{2006ApJ...651.1130G}. An alternative scenario would also be 
conceivable, in which the CO emission could be permanent, arising from a circumstellar ring or shell 
and being detectable only during phases in which no photospheric absorption compensates the emission 
\cite{2019MNRAS.483.3792K}. Support for the latter scenario is provided by the fact that the 
CO emission features remain at constant radial velocities (along with other emission lines 
formed in the circumstellar environment such as [Ca\,{\sc ii}] and Fe\,{\sc i}, 
\cite{2019MNRAS.483.3792K}) whereas the absorption components change from red- to blue-shifted, 
in phase with the pulsation cycle.

The object $\rho$~Cas is, to date, the best monitored YHG in the NIR. For many other YHGs 
observations in the $K$-band have been taken only sporadically. Hence, not much can be said 
or concluded about the variability rhythm of their CO band features. Occasional CO band 
emission has been reported for the galactic objects HD~179821 \cite{1994ApJ...420..783H, 
1995A&A...299...69O}, V509~Cas \cite{1981ApJ...248..638L}, [FMR2006]~15 
\cite{2008ApJ...676.1016D}, and the two LMC objects HD~269723 and HD~269953 
\cite{1988ApJ...334..639M, 2013A&A...558A..17O}. The latter even displays emission from hot 
water vapor and is, so far, the only evolved massive star with such emission features from its 
environment \cite{2022BAAA...63...65K}.

\subsection{Motivation and Aims}

The appearance of CO band emission in the NIR spectra of B[e]SGs and YHGs suggests that 
similar physical conditions may prevail in the circumstellar environments of both groups
of objects. In depth studies of these conditions are rare though. 

For the B[e]SGs in the Magellanic Clouds CO, column densities, temperatures, and 
$^{12}$CO/$^{13}$CO \textls[-25]{isotopic ratios were determined based on medium-resolution $K$-band} 
\mbox{spectra \cite{2013A&A...558A..17O}.} The resolution of these spectra was, however, too low to 
derive the gas kinematics with high confidence. On the other hand, high-resolution $K$-band 
spectra of the Galactic B[e]SG sample allowed to obtain the CO dynamics 
\cite{2012BAAA...55..123M, 2018MNRAS.480..320M}, whereas in most cases the spectral coverage 
was too short to infer about the density and temperature of their hot molecular environment.

%\unskip

The situation for the YHGs is even worse. Besides for $\rho$~Cas \cite{2006ApJ...651.1130G} 
and for the LMC object HD\,269953 \cite{2013A&A...558A..17O, 2022BAAA...63...65K} no attempts
have been undertaken so far to study their warm molecular environments in more detail and to 
derive the parameters of the CO band emitting regions.

\begin{figure}[H]
%\centering
\includegraphics[width=\textwidth]{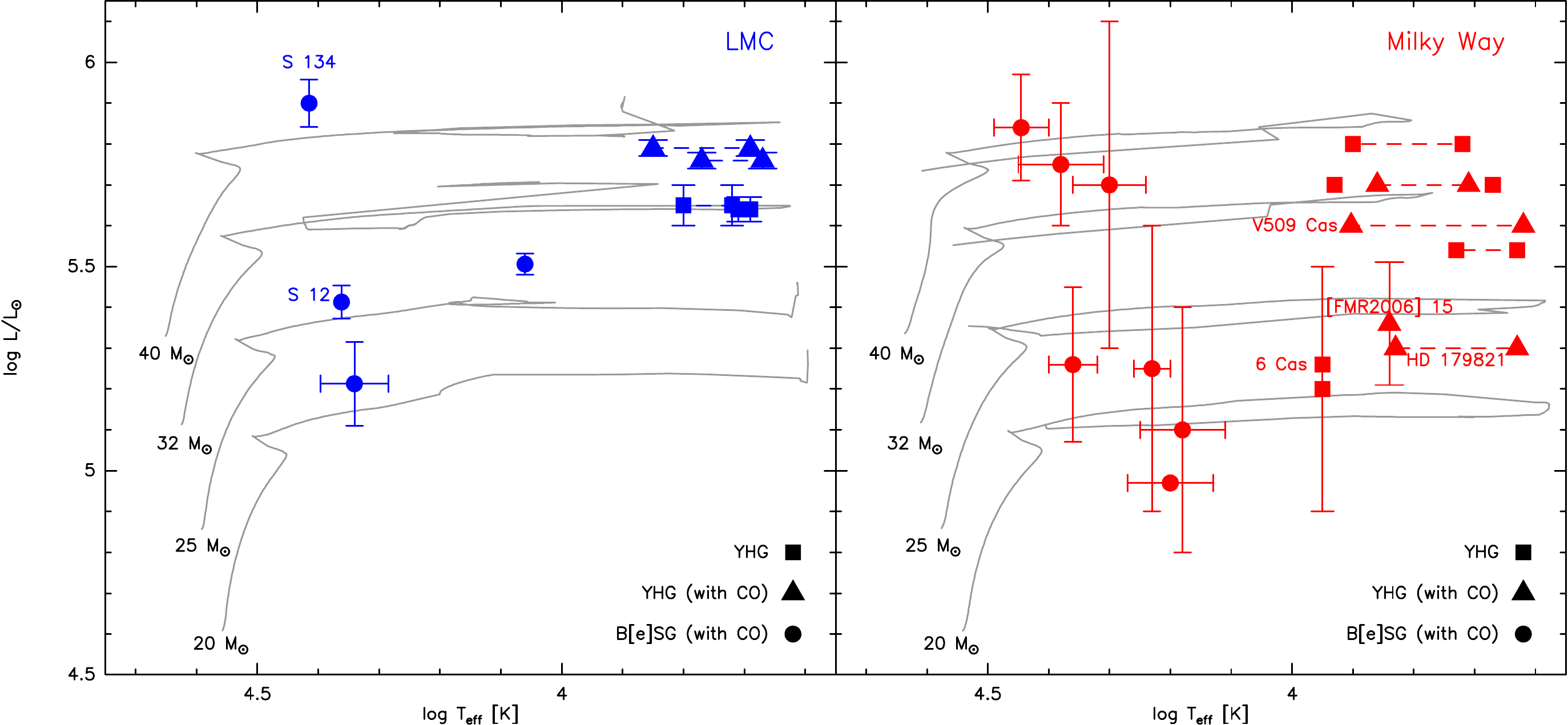}
\caption{{HR diagram with evolutionary tracks} for {LMC ($Z = 0.006$, \cite{2021A&A...652A.137E}, left panel) and solar ($Z=0.014$, \cite{2012A&A...537A.146E}, right panel)} 
metallicity for models of rotating stars {($v/v_{\rm crit} = 0.4$)} with initial masses from $20-40$\,M$_{\odot}$. Shown are the 
positions of  {LMC} (blue symbols) {and Galactic (red symbols)} objects. Only B[e]SGs 
with hot circumstellar molecular CO gas are shown. These populate  {similar} evolutionary tracks as the 
YHGs. The minimum and maximum temperature values (where known) of the YHGs are connected by dashed 
lines. YHGs with reported (at least once) CO band emission are shown with triangles. The stars 
of the current study are labeled.
\label{fig:HRD}}
\end{figure}

Therefore, we started to systematically observe both the B[e]SGs and the YHGs in the Milky Way 
and Magellanic Clouds to fill this knowledge gap. A further motivation for our research is 
provided by the location of the B[e]SGs with CO band emission in the HR diagram. As has been
mentioned previously for the Magellanic Clouds sample \cite{2019Galax...7...83K}, these objects 
cluster around luminosity values $\log L/L_{\odot} \simeq 5.0-{5.9}$, whereas B[e]SGs that are 
more luminous  {than $\sim$5.9} or less luminous than {$\sim$}5.0 do not 
show CO band emission. The same holds when  {inspecting} the Galactic B[e]SGs. 
{In the left panel of Figure~\ref{fig:HRD} we depict the LMC objects, whereas in the right panel  
we display the Galactic sample}\endnote{We excluded 
two Galactic B[e]SGs from this plot. With a literature luminosity value of $\log L/L_{\odot} = 4.33\pm 
0.09$ \cite{2011A&A...526A.107M}, the luminosity
 of HD~62623 is considerably lower than for the other 
B[e]SGs with CO bands. However, its distance with a parallax value of $0.59\pm 0.17$ is not well 
constrained. The object HD~327083 turned out to be misclassified and has been removed from the B[e]SGs 
list (Cidale et al., in preparation).} from \cite{2018MNRAS.480..320M,2020MNRAS.493.4308K}. 
The YHGs in the {LMC} 
 \cite{2022MNRAS.511.4360K} and the Milky Way \cite{1998A&ARv...8..145D} 
have been added to the plots\endnote{We omit the SMC objects, because we do not have 
new data for any of the SMC B[e]SGs, and there are currently no confirmed YHGs in the SMC.}. Their 
positions with maximum and minimum effective temperature values, as reported in the literature, have 
been connected with dashed lines. {Errors in temperature and luminosity are indicated when 
provided by the corresponding studies.} 

Interestingly, the YHGs and the B[e]SGs with CO band emission share  {similar} 
evolutionary tracks, {which is particularly evident for the Galactic objects}. This raises the 
question of whether evolved stars in this {particular} mass range suffer from specific instabilities in the 
blue and yellow temperature regimes independent of their evolutionary state. Such instabilities need 
to have the potential to drive mass ejections or eruptions, and the released mass would have to be 
dense and cool enough to create the required conditions for the formation of significant amounts of 
molecules generating intense band emission.

\subsection{Selection of Targets}

For our {current} study, we have selected two B[e]SGs from the LMC. These are the objects 
LHA~120-S~12 and LHA~120-S~134. Both are known to display CO band emission (see, e.g., 
\cite{2013A&A...558A..17O}), but both of them lack high-resolution $K$-band spectra to derive 
the CO kinematics. Furthermore, we have selected four Galactic YHGs, [FMR2006]~15, HD~179821, 
V509~Cas, and 6~Cas. Of these, only the former three objects were reported in the literature to 
display (at some epochs) CO band emission. The basic stellar parameters of all objects, as obtained 
from the literature, are listed in Table~\ref{tab:stelparam}.

\begin{table}[H]
\caption{Stellar parameters. {Errors are given where available.}}\label{tab:stelparam}
%\centering
%\tablesize{\footnotesize} %% You can specify the fontsize here, e.g., \tablesize{\footnotesize}. If commented out \small will be used.
\footnotesize
\setlength{\tabcolsep}{3.5mm}
\begin{tabular}{lrrccrrr}
\toprule
\textbf{Object}	& \boldmath{$G$} & \boldmath{$K_{s}$} & \boldmath{$\log T_{\rm eff}$} & 
   \boldmath{$\log L/L_{\odot}$} &  \textbf{Ref.} & \boldmath{$d$} & \textbf{Ref.} \\
 & \textbf{[mag]} & \textbf{[mag]} & \textbf{[K]} & &   & \textbf{[kpc]} &  \\
\midrule
LHA~120-S~12    & $12.4$ &  $10.2$ & $4.36$ & $5.34\pm0.04$ &  \protect{\cite{1986A&A...163..119Z}} & $49.6\pm0.5$ & \protect{\cite{2019Natur.567..200P}}\\
LHA~120-S~134   & $11.4$ &  $8.6$  & $4.41$ & $5.90\pm0.06$ &  \protect{\cite{1986A&A...163..119Z}} & $49.6\pm0.5$ & \protect{\cite{2019Natur.567..200P}}\\
\midrule
$ $[FMR2006]~15 & $19.5$ &  $6.7$  & $3.84$ & $5.36\pm0.15$ &  \protect{\cite{2008ApJ...676.1016D}} & $6.6\pm0.9$  & \protect{\cite{2008ApJ...676.1016D}}\\
6~Cas           &  ---   &  $3.4$  & $3.93$ & $5.13$ &  \protect{\cite{1999A&A...350..970K}} & $2.8\pm0.3$ & \protect{\cite{2021A&A...646A..11M}}\\
V509~Cas        &  $5.0$ &  $1.7$  & $3.90$ & $5.60$ &  \protect{\cite{2017ASPC..508..239A, 1997MNRAS.290L..50D}} &  $1.4\pm 0.5$   & \protect{\cite{2012A&A...546A.105N}} \textsuperscript{a}\\
HD~179821       &  $7.5$ &  $4.7$  & $3.83$ & $5.30$ &  \protect{\cite{2008BaltA..17...87K, 1998A&ARv...8..145D}} & $5.3\pm 0.3$  & \protect{\cite{2008BaltA..17...87K, 2020yCat.1350....0G}}\\
\bottomrule
\end{tabular}%\\
%\begin{tabular}{@{}c@{}} 
\\
\noindent{\footnotesize{Note: The $G$ and $K_{s}$-band magnitudes are from \textsl{GAIA} Early 
Data Release 3 {\cite{2020yCat.1350....0G}} and from the 2MASS point source catalog \cite{2003yCat.2246....0C}, respectively. The YHG effective temperatures refer to the hot state. No \textsl{GAIA} $G$-band measurement is available for 6~Cas.  
  \textsuperscript{a}~Listed distance is based on the \textsl{HIPPARCOS} parallax of $0.73\pm 0.25$. The new \textsl{GAIA} Early 
Data Release 3 parallax of $0.2507\pm 0.0633$ \cite{2020yCat.1350....0G} places the object at a distance of $\sim$4 kpc.  }}%\\
%\end{tabular}
\end{table}

%%%%%%%%%%%%%%%%%%%%%%%%%%%%%%%%%%%%%%%%%%
\section{Observations and Data Reduction}

High-resolution spectra ($R\sim$ 50,000) of the two B[e]SGs were acquired with the visitor 
spectrograph Phoenix \cite{2003SPIE.4834..353H} mounted at GEMINI-South. The spectra were taken on 
\mbox{20 December 2004} and 30 November 2017 under program IDs GS-2004B-Q-54 and GS-2017B-Q-32. The 
observations were carried out in the $K$-band with two different filters, K4396 and K4308. The 
central wavelength was chosen such that the wavelength ranges cover the first and second band heads of 
the first-overtone CO band emission. 

Medium-resolution $K$-band spectra of the YHGs have been acquired with the Gemini 
Near-InfraRed Spectrograph (GNIRS, \cite{2006SPIE.6269E..14E, 2006SPIE.6269E..4CE}) at 
GEMINI-North under Program IDs GN-2019A-Q-204, GN-2019B-Q-418, and GN-2021A-Q-315. 

The spectrum of [FMR2006]~15 was observed on 12 May 2019 centered on $\lambda = 2.35\,\upmu$m.
The instrument configuration was a short camera (0.15'' per pixel) with the 0.3'' slit and the 
111\,l\,mm$^{-1}$ grating, resulting in a resolving power of $R\sim$ 5900.
 
V509 Cas and 6 Cas were observed on 21 December 2019 with the instrumental configuration: Long 
camera (0.05'' per pixel) with the 0.10'' slit and the 32\,l\,mm$^{-1}$ grating which provides a 
resolving power of $R\sim 5100$. The observations were centered on $\lambda = 2.35\,\upmu$m.

HD 179821 was observed on 7 April 2021 with two different central wavelengths $\lambda = 
2.14\,\upmu$m and $2.33\,\upmu$m and with the following instrument configuration: a short camera 
(0.15'' per pixel), the 0.3'' slit, and the 111\,l\,mm$^{-1}$ grating, resulting in a resolving 
power of $R\sim 5900$.

For all objects, a telluric standard star (usually a late B-type main sequence star) was 
observed close in time and airmass. For optimal sky subtraction, the star was positioned at 
\textls[-25]{two different locations along the slit (A and B), and the observations were carried out in 
ABBA cycles.} \textls[-25]{Data reduction and telluric correction were performed using standard 
IRAF}\endnote{IRAF is distributed by the National Optical Astronomy Observatory, which is 
operated by the Association of Universities for Research in Astronomy (AURA) under cooperative 
agreement with the National Science Foundation.} tasks. The reduction steps consist of 
subtraction of AB pairs, flat-fielding, wavelength calibration (using the telluric lines), and 
telluric correction. The observing log is given in Table~\ref{tab:obs}, where we list the star
name, object class, observing date (UT), used instrument, covered wavelength 
range, spectral resolution $R$, and resulting signal-to-noise ratio (SNR).

\begin{table}[H]
\caption{Observation log.}
%\centering
%% \tablesize{} %% You can specify the fontsize here, e.g., \tablesize{\footnotesize}. If commented out \small will be used.
\small
\setlength{\tabcolsep}{2.7mm}
\begin{tabular}{lcccccr}
\toprule
\textbf{Object}	& \textbf{Class}	&\textbf{Obs Date} & \textbf{Instrument} & \boldmath{$\lambda_{\rm min} - \lambda_{\rm max}$} & \boldmath{\textbf{R}} & 
\textbf{SNR}\\
 & & \textbf{[yyyy-mm-dd]} & & \boldmath{[$\upmu$\textbf{m}]} & & 
 \\
\midrule
LHA 120-S 12    & B[e]SG &  2004-12-20   & Phoenix &  2.291--2.300 & 50,000 &  35\\
                &        &  2017-11-30   & Phoenix &  2.319--2.330 & 50,000 &  20\\
LHA 120-S 134   & B[e]SG &  2017-11-30   & Phoenix &  2.290--2.299 & 50,000 &  40\\
                &        &  2017-11-30   & Phoenix &  2.319--2.330 & 50,000 & 100\\
\midrule
$ $[FMR2006] 15 &  YHG   &  2019-05-12   &  GNIRS  &  2.257--2.440 &  5900 & 300\\
6 Cas           &  YHG   &  2019-12-21   &  GNIRS  &  2.232--2.453 &  5100 & 100\\
V509 Cas        &  YHG   &  2019-12-21   &  GNIRS  &  2.233--2.453 &  5100 & 140\\
HD 179821       &  YHG   &  2021-04-07   &  GNIRS  &  2.046--2.424 &  5900 & 250\\
\bottomrule
\label{tab:obs}
\end{tabular}
\end{table}

%%%%%%%%%%%%%%%%%%%%%%%%%%%%%%%%%%%%%%%%%%
\section{Results}

\subsection{Description of the Spectra}

\begin{figure}[H]
\centering
\includegraphics[width=\textwidth]{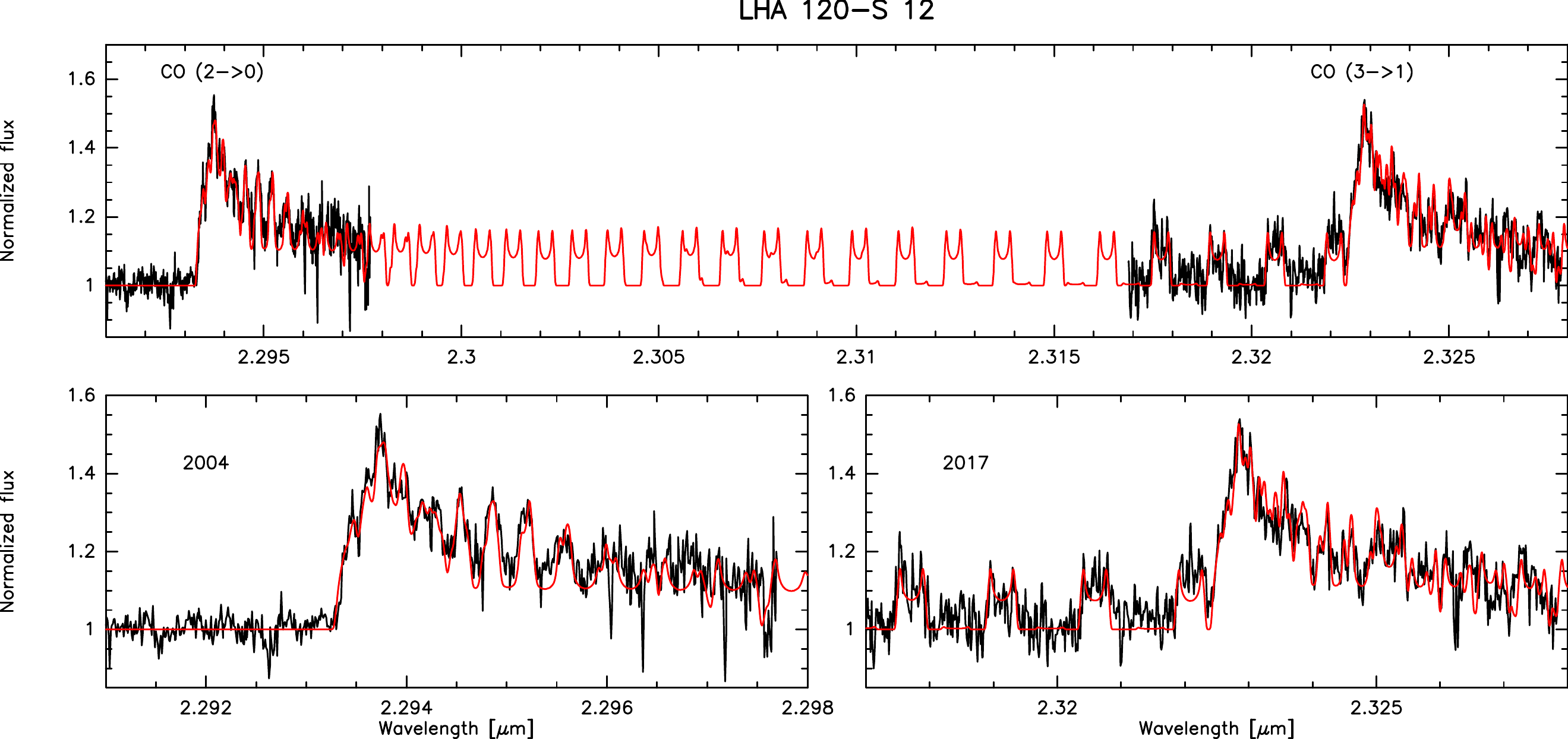}
\smallskip

\includegraphics[width=\textwidth]{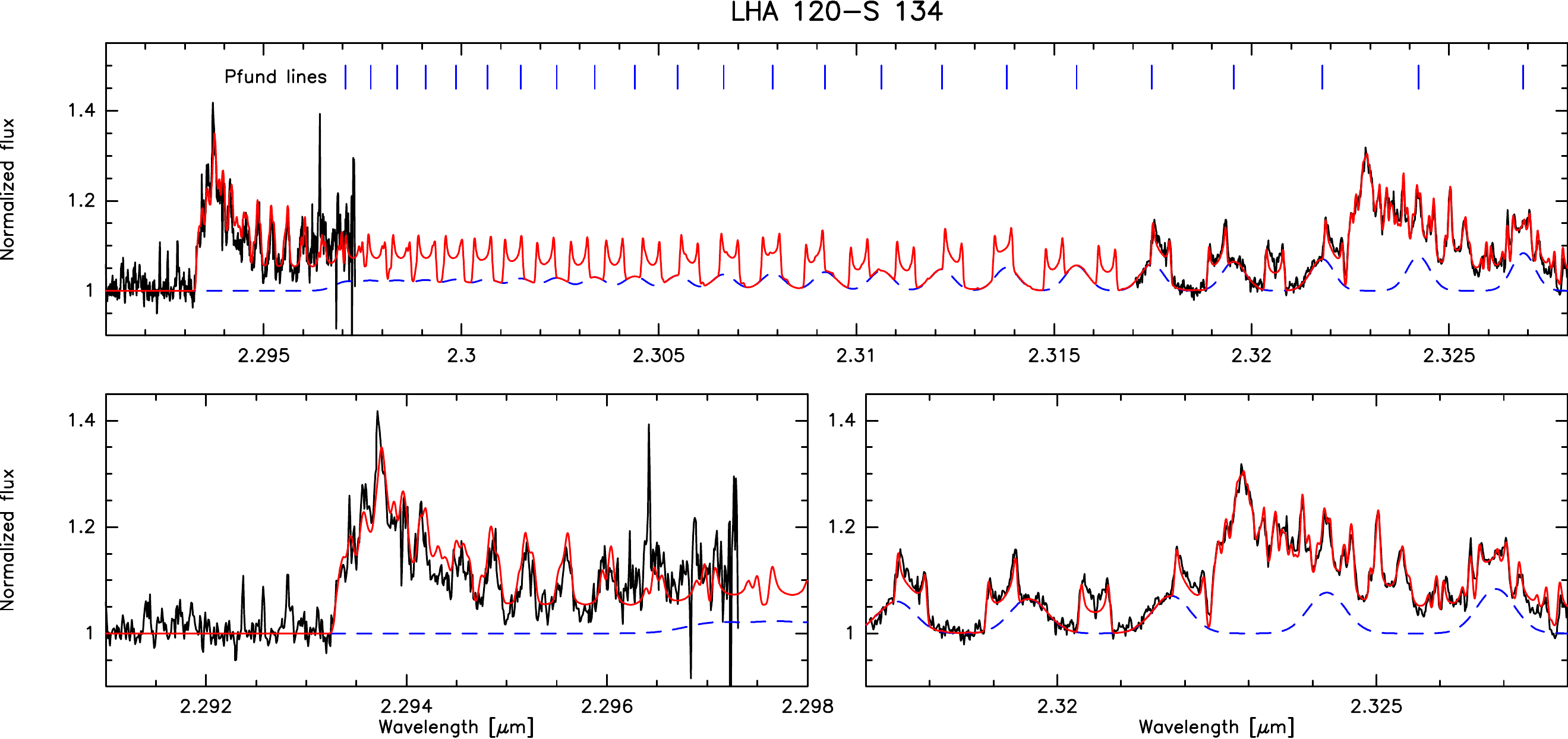}
\caption{Best fitting model (red) to the normalized Phoenix spectra (black) of LHA 120-S 12 
(\textbf{top}) and LHA 120-S 134 (\textbf{bottom}). For each star we display the entire fit to the total 
spectrum (top panels) and the zoom to the band heads (bottom panels). Emission from the Pfund 
series (blue dashed line), detected in the spectrum of LHA 120-S 134, is included in the total 
fit. 
\label{fig:fits-Phoenix}}
\end{figure}

CO band head emission is detected in both B[e]SGs (black lines in 
Figure~\ref{fig:fits-Phoenix}) despite the low quality of some of the spectral pieces and some 
telluric remnants in the red parts of the short wavelength portions (left lower panels). The 
spectrum of LHA 120-S 134 contains additional emission lines from the hydrogen Pfund series. 
Intense emission in these recombination lines has already been reported from that star based on 
medium-resolution spectra \cite{2013A&A...558A..17O}. No contribution from the Pfund lines is 
seen in the spectra of LHA 120-S 12.

%\unskip

\begin{figure}[H]
%\centering
\includegraphics[width=\textwidth]{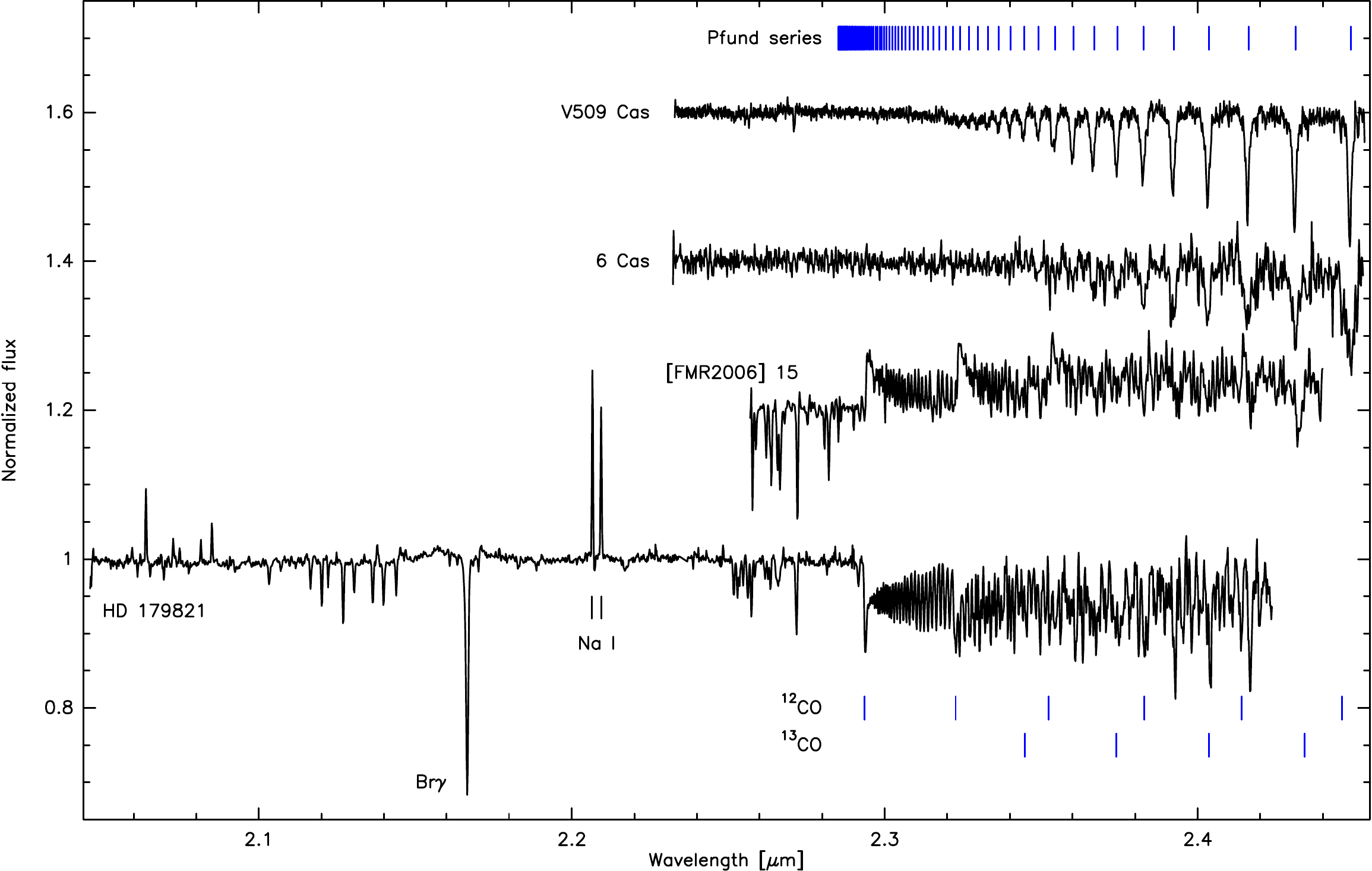}
\caption{Normalized medium-resolution $K$-band spectra of the YHGs taken with GNIRS. For better 
visualization, the spectra have been offset along the flux axis. Positions of the CO band heads 
and of the lines from the Pfund series are marked by ticks. The lines of Br$\gamma$ and of the 
Na\,{\sc i} doublet are labeled as well.
\label{fig:obs-GNIRS}}
\end{figure}

The $K$-band spectra of the YHGs display a diverse appearance, as depicted in 
Figure~\ref{fig:obs-GNIRS}. Two stars possess just Pfund lines in absorption (V509~Cas and 
6~Cas) and an otherwise featureless spectrum, in agreement with their high effective 
temperature ($T_{\rm eff} \geq 8000$\,K, see Table~\ref{tab:stelparam}). The star [FMR2006]~15 shows 
CO band emission overimposed on the atmospheric spectrum of a presumably late-type star. In 
HD~179821 the CO bands and the Br\,$\gamma$ line are in absorption along with numerous other 
photospheric lines, whereas the Na\,{\sc i} $\lambda\lambda$2.206,2.209 doublet shows prominent 
emission. 

%\unskip

\subsection{Modeling of the CO Band and Pfund Line Emission}

We model the CO emission using our molecular disk code \cite{2000A&A...362..158K} that has 
been developed to compute the ro-vibrational bands from a rotating ring (or disk) of 
circumstellar gas under local thermodynamic equilibrium conditions. 
The calculations are carried out for the two main isotopes, $^{12}$CO and $^{13}$CO 
\cite{2009A&A...494..253K, 2013A&A...558A..17O}. 

The high-resolution spectra of the two B[e]SGs display kinematic broadening of the individual 
ro-vibrational lines in the form of a double-peaked profile (Figure~\ref{fig:fits-Phoenix}). Such 
a profile can be interpreted either as rotation around the central object or as an equatorial 
outflow. Since B[e]SGs are known to be surrounded by (quasi-)Keplerian rotating disks, the 
assumption of rotation as the most likely broadening mechanism seems to be justified. 
The situation for the YHG star [FMR2006]~15 is less clear. The medium-resolution spectrum 
provides no clear hint about a possible double-peaked shape of the individual lines. Therefore, 
we can only derive an upper limit for a possible rotational (or outflow) contribution to the 
\linebreak total dynamics.

Due to the high sensitivity of the CO band intensity to the gas temperature and column 
density, the observed emission traces the hottest and densest molecular regions. Therefore, it 
is usually sufficient to consider a single ring of gas with constant density and temperature, 
reducing the number of free parameters to the column density $N_{\rm CO}$, temperature $T_{\rm 
CO}$, the isotope ratio $^{12}$CO/$^{13}$CO, and the gas kinematics split into contributions of 
the rotation velocity projected to the line of sight $v_{\rm rot,los}$, and of a combined 
thermal and turbulent velocity in the form of a Gaussian component $v_{\rm Gauss}$.

The best-fitting CO parameters obtained for the three objects are listed in 
Table~\ref{tab:CO}, and the total emission spectra are included (in red) in 
Figure~\ref{fig:fits-Phoenix} for the two B[e]SGs. It is noteworthy that the two CO band heads
of LHA~120-S~12 can be reproduced fairly well with the same model parameters, despite the fact 
that the spectral pieces have been observed \mbox{13 years} apart. This implies that the ring of CO 
gas around LHA~120-S~12 is stable on longer timescales. 

\begin{table}[H]
\caption{Best fitting CO parameters.}\label{tab:CO}
%\centering
%% \tablesize{} %% You can specify the fontsize here, e.g., \tablesize{\footnotesize}. If commented out \small will be used.
\small
 \setlength{\tabcolsep}{3.5mm}
\begin{tabular}{lccccc}
\toprule
\textbf{Object}	& \boldmath{$T_{\rm CO}$}	& \boldmath{$N_{\rm CO}$} & \boldmath{$v_{\rm rot, los}$} & \boldmath{$v_{\rm Gauss}$} & \textbf{$^{12}$CO/$^{13}$CO}\\
 & \textbf{[K]} & \boldmath{[$\times 10^{21}$ \textbf{cm}$^{-2}$]} &  \textbf{[km\,s$^{-1}$]} & \textbf{[km\,s$^{-1}$]} & \\
\midrule
LHA~120-S~12    & $2800\pm 200$ &  $1.5\pm 0.2$   & $27\pm 2$ & $3\pm 1$  & $20\pm 2$\\
LHA~120-S~134   & $2300\pm 100$ &  $2.0\pm 0.1$   & $30\pm 1$ & $1.5\pm 0.5$  &  $15\pm 2$\\
\midrule
$ $[FMR2006]~15  & $3000\pm 200$ &  $2.0\pm 0.2$   & $20\pm 5$ & $15\pm 5$ & $4\pm 2$ \\
{$ $[FMR2006]~15}  & {$3000\pm 200$} &  {$2.0\pm 0.2$}   & {$0$} & {$20\pm 5$} & {$4\pm 2$} \\
\bottomrule
\end{tabular}%\\
%\begin{tabular}{@{}c@{}} 
\\
\noindent{\footnotesize{Note: The $^{12}$CO/$^{13}$CO values for LHA~120-S~12 
and LHA~120-S~134 have been derived by \cite{2010MNRAS.408L...6L,2013A&A...558A..17O}, respectively. Our Phoenix spectra do not reach the wavelength 
region of the $^{13}$CO bands.
}}
%\end{tabular}
\end{table}

The spectrum of LHA~120-S~134 displays emission from the hydrogen Pfund line series 
superimposed on the CO band spectrum\endnote{Pfund line emission is also reported from 
LHA~120-S~12 \cite{2010MNRAS.408L...6L}, but in that star the maximum detected Pfund 
transition is with Pf(31) arising at 2.34\,$\mu$m clearly outside our spectral coverage.}. 
These Pfund lines appear to be broad with no indication of a double-peaked profile shape. 
 To include the 
contribution of these lines to the total emission spectrum, we apply our code developed for the 
computation of the hydrogen series according to Menzel case B recombination, assuming that 
the lines are optically thin \cite{2000A&A...362..158K}. We fix the electron temperature at 
10,000\,K, which is a reasonable value for  ionized  {gas} around an 
OB supergiant star and,  using a Gaussian profile, 
{we obtain} a velocity of $53\pm 3$\,km\,s$^{-1}$. {Similar velocity values for   
the lines from the Pfund series have been found for various B[e]SGs (see \cite{2010MNRAS.408L...6L,
2013A&A...558A..17O, 2015AJ....149...13M}). These rather low values
compared with the wind velocities of classical B supergiants might suggest that the Pfund lines form
in a wind emanating from the surface of the ionized part of the circumstellar disk.} The 
electron density can be derived from the maximum number of {the Pfund} %Kraus: The article disappeared from the previous text "of the series", which is not correct. To avoid misunderstandings we specified "of the Pfund series".
series visible in the spectrum. 
For LHA~120-S~134, this number corresponds to the line Pf(57), resulting in an electron density 
of $(5.8\pm 0.5)\times 10^{12}$\,cm$^{-3}$ within the Pfund line forming region. Having the 
parameters for the Pfund emission fixed, we compute the contribution of the Pfund series to the 
total emission spectrum of LHA~120-S~134. This contribution is shown in blue in 
Figure~\ref{fig:fits-Phoenix}.

%\unskip

\begin{figure}[H]
%\centering
\includegraphics[width=\textwidth]{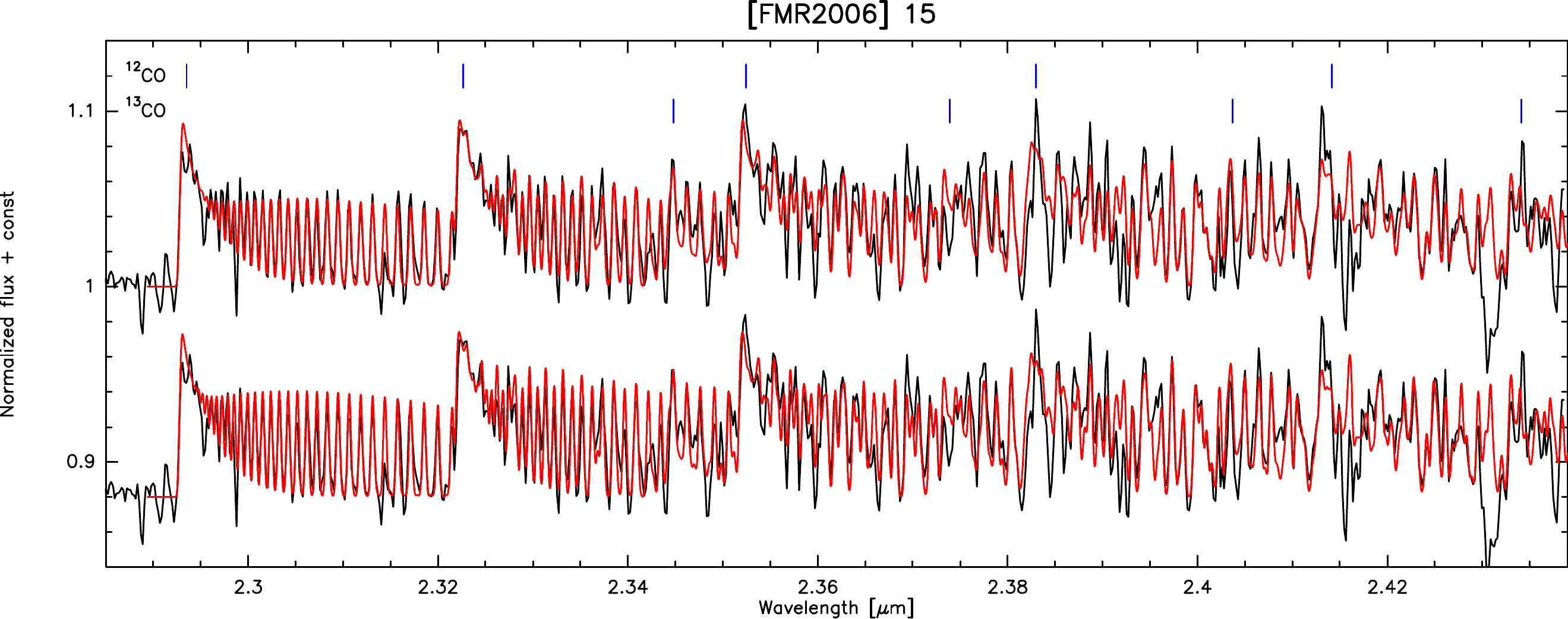}
\caption{Best fitting model (red) to the observed (black) $K$-band spectrum of [FMR2006]~15
{for the model including a rotational component (\textbf{top}) and a pure Gaussian broadening (\textbf{bottom})}.
\label{fig:FMR-CO}}
\end{figure}

The best-fitting CO model for the YHG [FMR2006]~15 is depicted in Figure~\ref{fig:FMR-CO} 
{(top)}. It should be noted that the contribution of the rotation (respectively outflow) 
component should be considered an upper limit. A double-peaked profile corresponding to such a 
velocity might be hidden within the CO band structure. Only high-resolution observations will tell 
whether such a profile component is really included.  {We found} that 
 {a model} omitting this velocity component and using instead just a single 
Gaussian profile {(added to Table\,\ref{tab:CO})}  {results in a similar but slightly 
less satisfactory fit because the intensity of many individual ro-vibrational lines is overestimated 
in the short wavelength domain (Figure~\ref{fig:FMR-CO}, bottom).}

The rotation velocity of the CO gas allows us to estimate the distances of the CO-emitting rings
from the central object. For this, we need to know the current stellar masses. Considering the 
stellar effective temperatures, luminosities (from Table\,\ref{tab:stelparam}), and the carbon isotope 
ratios (from Table\,\ref{tab:CO}), we search for the best matching evolutionary models for each of our 
targets utilizing the stellar evolution track interpolation tool 
SYCLIST\endnote{\url{https://www.unige.ch/sciences/astro/evolution/en/database/syclist/}}.
%(accessed on 2023 June 3)
To reproduce the observed carbon isotope ratios, we vary the initial stellar rotation rate
between zero and the maximum offered value of 0.4 and select those evolutionary tracks that
fit both the stellar location in the HRD and the carbon isotope ratio. The metallicities used are 
0.006 for the LMC objects and 0.014 for the Galactic star. Table\,\ref{tab:masses} lists 
the stellar radii of our targets along with the most likely initial masses, initial rotation rate, and 
the resulting current masses. We note that according to these evolutionary tracks, the B[e]SGs seem to 
have evolved just off the main sequence, whereas the low carbon isotope ratio measured in [FMR2006]~15 
clearly places this object on the post-RSG path. For the reason that no proper values for the disk inclination 
angles are known, we provide the distances of the CO-emitting rings as a function of the inclination 
angle. These are lower limits to the real distances. We refrain from adding a distance of the CO 
emitting region for [FMR2006]~15 because of the speculative nature of its rotation component.

\begin{table}[H]
\caption{{Estimated current stellar masses and distances of the CO emitting rings from the star.}}\label{tab:masses}
%\centering
%% \tablesize{} %% You can specify the fontsize here, e.g., \tablesize{\footnotesize}. If commented out \small will be used.
 \setlength{\tabcolsep}{2.8mm}
\begin{tabular}{lrrcrcc}
\toprule
\textbf{Object}	& \boldmath{$R_{*}$}	& \boldmath{$M_{\rm in}$} & \boldmath{$v/v_{\rm crit}$} & \boldmath{$M_{*}$} & \boldmath{$r_{\rm CO}/(\sin i)^{2}$} & \boldmath{$r_{\rm CO}/(\sin i)^{2}$}\\
 & \boldmath{[$R_{\odot}$]} & \boldmath{[$M_{\odot}$]} &  \textbf{[km\,s$^{-1}$]} & \boldmath{[$M_{\odot}$]} & \textbf{[cm]} &  \boldmath{[$R_{*}$]} \\
\midrule
LHA~120-S~12     & $30$  &  $26$  & $0.25$ & $25$  & $4.6\times 10^{14}$ & $218.0$ \\
LHA~120-S~134    & $44$  &  $45$  & $0.36$ & $40$  & $5.9\times 10^{14}$ & $192.7$ \\
\midrule
$ $[FMR2006]~15  & $333$ &  $25$  & $0.40$ & $13$  & --- & --- \\
\bottomrule
%\end{tabular}\\
%\begin{tabular}{@{}c@{}} 
%\multicolumn{1}{p{\textwidth -.68in}}{{Note:} We refrain from adding a CO distance for [FMR2006]~15. 
%If to rotational component will be confirmed to be real, the CO gas might not be revolving [FMR2006]~15 itself but a hidden companion as 
%discussed in Section\,\ref{sect:discussion}.
%}%\\
\end{tabular}
\end{table}

%%%%%%%%%%%%%%%%%%%%%%%%%%%%%%%%%%%%%%%%%%
\section{Discussion}\label{sect:discussion}

We have detected CO band emission in three objects and CO band absorption in one object in our sample.
To assess a possible CO variability, we summarize in Table~\ref{tab:CO-obs} 
information about previous detections of CO band features in the $K$-band from the literature, 
including our results.

The number of observations in the $K$-band over the last 3--4 decades is sparse for all 
objects, but the two B[e]SGs continuously display emission over a time interval of 32 years. 
The diversity in spectral resolution used for the observations makes it difficult to compare the 
shape and intensity of the emission bands and, hence, to judge variability. Only for
LHA\,120-S\,12, we can say that no significant variability seems to have taken place 
between our observations acquired in 2004 and in 2017.

For the YHGs, the situation is different. Three of them clearly show variability in their CO 
bands, ranging from emission over complete disappearance to absorption (or a combination of 
emission and absorption). The changes are on time scales ranging from months to years. The exception is 
6\,Cas, which has not been reported previously to have CO bands, and our own observations taken 
25 years later also lack any CO band features.

In the following, we briefly describe the known characteristics of each of our targets with 
respect to stellar variability and the properties of the circumstellar environments in order to 
incorporate our new observational results.

\begin{table}[H]
\caption{CO band detection in the $K$-band and CO variability.}\label{tab:CO-obs}
%\centering
%% \tablesize{} %% You can specify the fontsize here, e.g., \tablesize{\footnotesize}. If commented out \small will be used.
\small
 \setlength{\tabcolsep}{5.5mm}
\begin{tabular}{lcrcc}
\toprule
\textbf{Object}	& 	\textbf{Obs Date} & \boldmath{$R$} & \textbf{CO Bands} & \textbf{Ref.} \\
\midrule
LHA\,120-S\,12    & 1985-12-28
   &  450  & em & \protect{\cite{1988ApJ...334..639M}} \\
                & 2004-12-17   &  50,000 & em & TW \\
                
                & 2009-10-14   &  4500 & em & \protect{\cite{2010MNRAS.408L...6L, 2013A&A...558A..17O}} \\
                & 2017-11-30   &  50,000 & em & TW \smallskip \\
LHA\,120-S\,134   & 1985-12-29   &    450 & em & \protect{\cite{1988ApJ...334..639M}} \\
                & 2009-11-10   &   4500 & em & \protect{\cite{2013A&A...558A..17O}} \\
                & 2017-11-30   &  50,000 & em & TW \\
\midrule
$[$FMR2006]~15  &   2005-09-15   &  1000  & weak abs  & \protect{\cite{2006ApJ...643.1166F}} \\
                &   2006-05-05   &  17,000 & em & \protect{\cite{2008ApJ...676.1016D}} \\
                &   2006-08-12   &  17,000 & none & \protect{\cite{2008ApJ...676.1016D}} \\
                &   2019-05-12   &   5900 & em & TW \smallskip \\
6~Cas           &   1996-08-31   &   1800 & none & \protect{\cite{2002A&A...390.1033V}} \\
                &   2019-12-21   &   5100 & none & TW \smallskip \\
V509~Cas        &   1979-1980    &  32,000  & em + abs (variable) & \protect{\cite{1981ApJ...248..638L}} \\
                &   1988   &   N.A. & none &  {\cite{2006ApJ...651.1130G}} \\
                &   2003-11-20   &   300 & none & {\cite{2006ApJ...651.1130G}}  \\
                &   2004-10-30   &   300 & none & {\cite{2006ApJ...651.1130G}}  \\
                &   2019-12-21   &   5100 & none & TW \smallskip \\
HD~179821       &   1989-07-14   &   1600 & em   &  \protect{\cite{1995A&A...299...69O}}\\
                &   1990-09-26   &    760 & em   &  \protect{\cite{1994ApJ...420..783H}}\\
                &   1991-11-04   &    330 & em   &  \protect{\cite{1994ApJ...420..783H}}\\
                &   1992 August  &   N.A. & none &  \protect{\cite{1995A&A...299...69O}}\\               
                &   1997-04-19   &   1800 & none & \protect{\cite{2002A&A...390.1033V}} \\
                &   2000-10-18   &   N.A. & none &  \protect{\cite{2006ApJ...651.1130G}}\\
                &   2021-04-07   &   5900 & abs  & TW \\
\bottomrule
\end{tabular}
\\
%\begin{tabular}{@{}c@{}} 
\noindent{\footnotesize{Note: em = emission; abs = absorption; TW = this work; N.A. = no information available.
}}%\\
%\end{tabular}
\end{table}

\paragraph{\bf LHA\,120-S\,12 (= SK\,$-$67\,23)}

The object
 LHA\,120-S\,12 was first mentioned in the catalog of H$\alpha$-emission stars in 
the Magellanic Clouds \cite{1956ApJS....2..315H}. Follow-up observations recorded an intense IR 
excess due to hot dust \cite{1984A&A...131L...5S}. The two-component wind associated with the 
star led to its classification as B[e]SG \cite{1986A&A...163..119Z}, and the high degree of 
intrinsic polarization suggested a high, but not fully edge-on, viewing angle towards the 
object and its dusty disk \cite{1992ApJ...398..286M}. Shell-line profiles were seen in the NIR,
consistent with the high inclination angle of the system \cite{2013A&A...558A..17O}. Observations 
with the
 \textit{Spitzer Space Telescope} revealed silicate dust within the circumstellar disk, 
based on weak features in the star's mid-IR spectrum, but no IR nebulosity has been detected in 
association with the object \cite{2010AJ....139.1993K}.

The first detection of CO band emission from LHA\,120-S\,12 was in 1985 based on a low-resolution
$K$-band spectrum \cite{1988ApJ...334..639M}. This spectrum clearly depicted four first-overtone 
band heads, similar to the observations collected in 2009. Modeling of the latter 
\cite{2010MNRAS.408L...6L} resulted in similar parameters for the CO temperature and column density 
 {to} those we found from the high-resolution spectra taken before (2004) and 
afterwards (2017). Moreover, we see no changes in the rotation velocity, projected to the line 
of sight, within the time span of 13 years between our two observations. Therefore, we believe 
that the CO-emitting ring revolving around LHA\,120-S\,12 is rather stable, {with no detectable outflow or inflow}.

\textls[-25]{A similar (projected) rotation velocity  {to that} for CO can be inferred from the 
double-peaked profiles of the [Ca{\sc ii}] lines resolved in high-resolution optical} spectra \mbox{of 
LHA\,120-S\,12 \cite{2012MNRAS.423..284A},} although a Gaussian component with a higher value 
might be necessary to smooth out the sharp synthetic double-peaked 
rotation profile in order to reproduce the shape of the observed forbidden lines.

\paragraph{\bf LHA\,120-S\,134 (= HD 38489, SK\,$-$69\,259, MWC~126)}

The object LHA\,120-S\,134 was listed in the catalog of B and A-type stars with bright hydrogen 
lines published in 1933 \cite{1933ApJ....78...87M} under the Mount Wilson Catalogue (MWC) 
number 126 with classification as Beq. Its hybrid spectra with narrow (equivalent widths of 
\mbox{30--50\,km\,s$^{-1}$)} emission lines of neutral and low-ionized metals in the optical spectral
\mbox{range \cite{1986A&A...163..119Z}} and very broad P~Cygni profiles (implying a wind terminal 
velocity of $\sim$2300\,km\,s$^{-1}$) of high-ionized metals in the ultraviolet
\cite{1983ApJ...273..177S}, along with the detected intense IR excess emission characteristic 
of hot circumstellar dust \cite{1984A&A...131L...5S}, resulted in the classification of the 
star as B[e]SG. A small inclination angle has been proposed for \mbox{LHA\,120-S\,134 
\cite{1986A&A...163..119Z},} and the relatively low measured degree of polarization supports a 
close to pole-on orientation of the star plus disk system \cite{1992ApJ...398..286M}.

\textls[-25]{LHA\,120-S\,134 is one of only two B[e]SGs showing broad emission from} \mbox{He\,{\sc ii} 
$\lambda$4686 \cite{1986A&A...163..119Z}.} The other object is LHA\,115-S\,18 in the SMC, which 
displays this (time variable) He\,{\sc ii} line in emission in 
concert with Raman-scattered O\,{\sc vi} emission and TiO molecular emission features 
\cite{2012MNRAS.427L..80T}. While for LHA\,115-S\,18, it was proposed that the peculiar 
spectral characteristics might point towards an LBV-like status of the star 
\cite{2012MNRAS.427L..80T}, a possible Wolf-Rayet companion was postulated to explain the 
spectral features of LHA\,120-S\,134 \cite{2014ApJ...788...83M}.
Although a solid proof for a Wolf-Rayet companion is still missing, LHA\,120-S\,134 has appeared
since then in the catalog of LMC Wolf-Rayet stars (WR~146).

The mid-IR spectrum of LHA\,120-S\,134, obtained with the \textit{Spitzer Space Telescope},
shows intense $10\,\upmu$m and weak $20\,\upmu$m emission features of amorphous silicate dust, and 
a faint and wispy nebulosity around the IR bright star was found with the telescope's imaging 
facilities \cite{2010AJ....139.1993K}. Follow-up optical imaging revealed that LHA\,120-S\,134
is located on the northeast rim of the superbubble of DEM L269 and on the western rim of the 
H\,{\sc ii} region SGS LMC-2 \cite{2021ApJS..252...21H}. Therefore, it is unclear if and how 
much of the optical and IR nebulosity might be related to LHA\,120-S\,134 itself.
 
In the NIR regime, the first mention of CO first-overtone emission dates back to 1985, when 
the star was observed with low-resolution \cite{1988ApJ...334..639M}. The next $K$-band 
spectrum was taken only 24 years later (see Table \ref{tab:CO-obs}). The new spectrum had 
higher spectral resolution, but the CO bands appeared to be similar to the previous detection. 
Our new, high-resolution spectrum was acquired after another time gap of 8 years. Our
modeling of the band spectrum revealed basically the same CO parameters (temperature and 
column density) as in 2009, with one addition, the projected rotational velocity. As for
LHA\,120-S\,12, this velocity is comparable to the one that might be inferred from the 
[Ca\,{\sc ii}] lines \cite{2012MNRAS.423..284A}, and we may conclude that also 
LHA\,120-S\,134 is surrounded by a stable, rotating ring of atomic and molecular gas.

\paragraph{\bf [FMR2006]~15 (= 2MASS J18375778-0652320)}

The star [FMR2006]~15 was recorded as object number 15 in a survey of the cool 
supergiant population of the massive young star cluster RSGC1 \cite{2006ApJ...643.1166F}. 
Based on the weak CO absorption detected in its $K$-band spectrum, a spectral type of 
G6\,I was allocated. In a follow-up investigation, it was proposed that [FMR2006]~15 is
most likely a YHG, based on the star's luminosity and spectral similarity to $\rho$\,Cas 
\cite{2008ApJ...676.1016D}. The star was assigned an effective temperature of $6850\pm 350$\,K 
and a luminosity of $\log L/L_{\odot} = 5.36^{+0.14}_{-0.16}$. With this temperature, the 
spectral type of [FMR2006]~15 is more likely G0 ($\pm2$ subtypes), and the luminosity implies 
that the star lies with about $M_{\rm ini} \sim 25$\,M$_{\odot}$ in the lower mass range of 
stars developing into YHGs (see Figure~\ref{fig:HRD}). This relatively low initial mass was 
considered to be the reason for the lack of detectable intense IR excess emission and of maser 
emission in contrast to the high-mass YHGs such as, e.g., IRC~+10420 
\cite{2008ApJ...676.1016D}.

The first low-resolution $K$-band spectrum of [FMR2006]~15 was taken in September 2005. At that time, 
very weak CO band absorption was seen \cite{2006ApJ...643.1166F}. Soon thereafter, the object was
re-observed twice with considerably higher spectral resolution. In May 2006, the spectrum displayed CO 
band emission and in August 2006, the emission had \mbox{disappeared \cite{2008ApJ...676.1016D}.} 
Our detection of intense CO band emission 13 years later is a clear indication that [FMR2006]~15 is 
embedded in circumstellar matter, even if the molecular gas is with 3000\,K much too hot for the 
condensation of dust grains. When modeling the CO bands from [FMR2006]~15, we noticed that the 
CO emission displays a blue-shift of $-77$\,km\,s$^{-1}$ with respect to the atmospheric 
absorption line spectrum. A similar behavior has been seen in IRC~+10420. This star displays 
blue-shifted emission in its IR hydrogen recombination lines \cite{1994A&A...281L..33O} and in 
its CO rotational transitions \cite{1996MNRAS.280.1062O}, which has been interpreted as 
emission formed in a close to pole-on seen bipolar outflow with the central star eclipsing the 
receding part of the emission. The outflow velocity from IRC~+10420 has been measured to be on 
the order of $\sim$40\,km\,s$^{-1}$, which is about half the value we measure for 
[FMR2006]~15. To date, nothing is known about a possible orientation of [FMR2006]~15, but we 
might speculate that a similar scenario as for IRC~+10420 might hold for that object as well. 
{In such a case, we would expect to have a Gaussian-like distribution of the velocity (see
bottom model in Figure~\ref{fig:FMR-CO}). 

On the other hand, if we consider the contribution from the rotation as real, the blue-shifted CO 
gas might revolve around a hidden companion, as in the case of the YHG HD~269953 in the LMC 
\cite{2022MNRAS.511.4360K}. In this object, the companion was proposed to be surrounded by a gaseous 
disk traced by numerous emission lines that display a time-variable radial velocity offset with 
respect to the photospheric lines of the YHG star. If [FMR2006]~15 is indeed a binary system, 
time-resolved K-band observations would be essential to derive the orbital parameters and to 
characterize the hidden companion.}

\begin{figure}[H]
%\centering
\includegraphics[width=\textwidth]{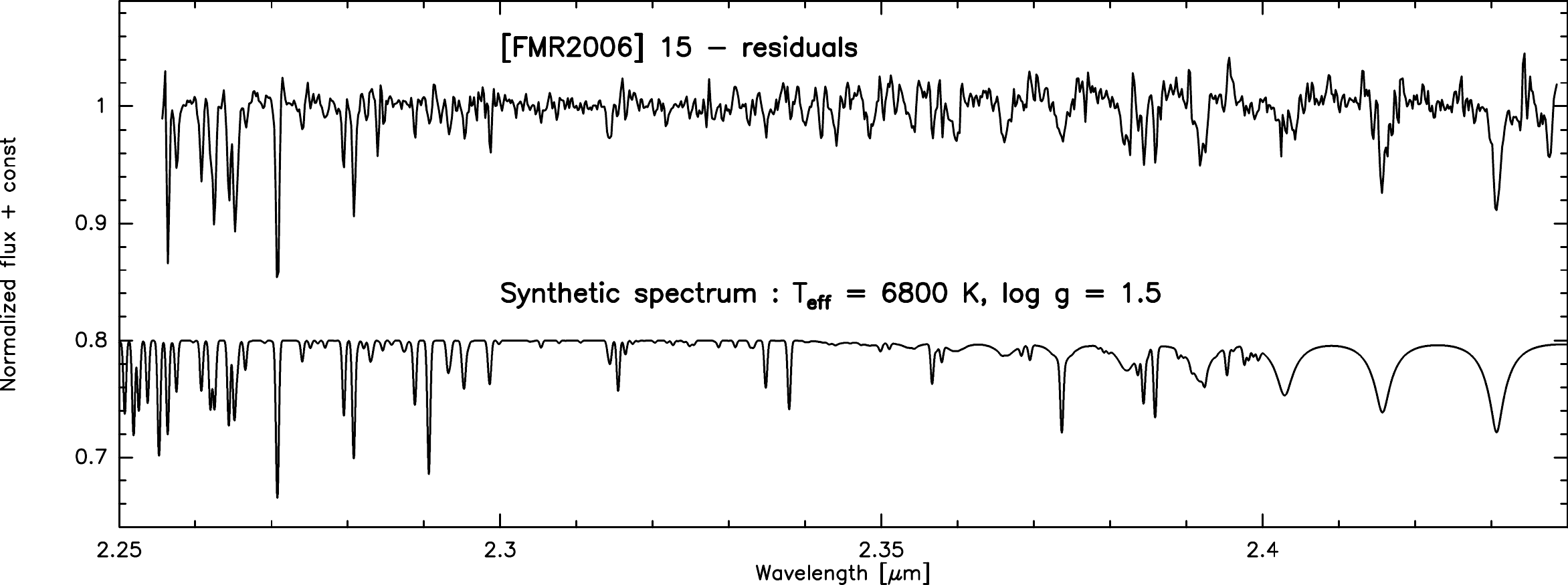}
\caption{Residual spectrum of [FMR2006]~15 after subtraction of the blue-shifted CO band 
emission. For illustration purposes, a synthetic model spectrum of a cool supergiant is shown 
as well (shifted down along the flux axis for better visualization) with parameters similar to those determined for 
[FMR2006]~15. \label{fig:FMR}}
\end{figure}

\textls[-25]{The  {best-fitting CO model} has been subtracted from the $K$-band spectrum of [FMR2006]~15,} and the 
residual is shown in Figure~\ref{fig:FMR}. Also included in this plot is a synthetic spectrum
of a cool supergiant star with $T_{\rm eff} = 6800$\,K and $\log g = 1.5$, similar to the 
values derived for [FMR2006]~15 \cite{2008ApJ...676.1016D}. This spectrum has been computed 
with the spectrum synthesis code Turbospectrum using MARCS atmospheric models 
\cite{2012ascl.soft05004P}. We note that the main photospheric features are decently well 
represented by this model. The short wavelength coverage of our spectrum and a possible 
alteration of the intrinsic stellar spectrum caused by the absorbing circumstellar gas 
impede a more decent classification of the object.

%\unskip

\paragraph{\bf 6~Cas (= V566~Cas, HR~9018, HD~223385)}

The object 6~Cas was reported to be a spectroscopic binary based on its composite spectrum in 
the ultraviolet \cite{1987A&A...181..300T}. While the main component is traced by the 
resonance lines of low-ionized metals with terminal wind velocities of about 330\,km\,s$^{-1}$, 
typical for A-type supergiants, the high-ionized resonance lines with P~Cygni profiles and 
terminal wind velocities of about 2400\,km\,s$^{-1}$ can be assigned to a significantly hotter 
O-type companion. Disentangling the spectra in the optical, to which the O star only very 
weakly contributes, resulted in the classification of the system as a A3 Ia (A) and a O9.5 II 
(B) component \cite{2021A&A...646A..11M}. The positions of the two stars are separated by 
about $1\farcs 5$ at a position angle of $\sim$195$^{\circ}$. Whether these two stars form 
indeed a bound binary system or whether they are just close in projection could not be solved 
yet. The small radial velocity variations detected for the A supergiant seem to be due to 
pulsations rather than orbital motion in a binary system \cite{2021A&A...646A..11M}.

The H$\alpha$ profile of 6~Cas resembles those seen in other hypergiants, such as 
HD~33579 and Schulte~12. It is superimposed on broad electron scattering wings 
\cite{2018ARep...62...19K}. Discrete absorption components (DACs) traveling through the broad 
absorption components of the wind profiles of H and Fe\,{\sc ii} lines were detected in 
6~Cas \cite{1995Ap&SS.232..217C}. Based on the atypical characteristics of a regular A-type 
supergiant, the star was assigned the status of a hypergiant.

We are aware of only one former $K$-band spectrum of 6~Cas, which has been taken with the 
ISO-SWS instrument \cite{2002A&A...390.1033V}. This spectrum shows no indication of CO bands, 
just the lines of the Pfund series in absorption, similar to our spectrum 
(Figure~\ref{fig:obs-GNIRS}). Such an otherwise featureless spectrum is consistent with 
a hot ($T_{\rm eff} > 8000$\,K) star, in agreement with its previous classification
(see Table\,\ref{tab:stelparam}).

\paragraph{\bf V509~Cas (= HR~8752, HD~217476)}

The star V509~Cas is one of the YHGs for which an outburst was recorded based on combined 
photometric and spectroscopic monitoring of the star \cite{2012A&A...546A.105N}. This outburst 
must have taken place around 1973, after a preceding 16-year-long period of reddening and 
cooling of the object from about 5000\,K down to about 4000\,K \cite{1975ApJ...202..743L}. 
Thereafter, the effective temperature of the star gradually increased until it reached a value 
of about 8000\,K around 2001 \cite{2012A&A...546A.105N}, where it has stabilized since then 
\cite{2017ASPC..508..239A}. During the ``heating'' period, several short-term drops in 
temperature were recorded, which were associated with phases of enhanced mass loss
\cite{2012A&A...546A.105N}.

Despite the outburst and the multiple mass-loss events, no extended nebulosity 
was detected at optical wavelengths so far \cite{2006AJ....131..603S}. Nevertheless,
the star must be embedded in an ionized circumstellar envelope traced by thermal emission
at radio wavelengths \cite{1977A&A....60..277S, 1978ApJ...220L.109H}. The ionization
of this envelope is most likely performed by the radiation field of a distant, hot 
main-sequence B1-type companion \cite{1978A&A....70L..53S}. The envelope is also the place 
where other emission lines are formed. Most prominent are the nebular lines of [N\,{\sc ii}] 
$\lambda\lambda$ \mbox{6548,6583 \cite{1965Obs....85...33S},} but also those of [O\,{\sc i}] 
$\lambda\lambda$ 6300,6364 and [Ca\,{\sc ii}] $\lambda\lambda$ 7291,7324 were identified 
\cite{2017ASPC..508..239A}, which were proposed to trace a possible  
Keplerian disk or ring based on the double-peaked profiles, particularly of the 
[Ca\,{\sc ii}] lines, for which a rotation velocity, projected to the line of sight, of 
about 40\,km\,s$^{-1}$ was derived \cite{2017ASPC..510..162A}. Support for such an 
interpretation comes from an optical spectroscopic monitoring of V509 Cas 
between 2015 and 2022. The observations revealed that the [Ca\,{\sc ii}] lines 
are stable in position and shape over the observing period of $\sim$7 years, in contrast to 
the photospheric absorption lines, whose shape and radial velocity are strongly influenced 
by the pulsation activity of the star \cite{2022not..confE..14K}. 

In the NIR, observations dating back to 1979--1980 detected CO features displaying both
emission and absorption components with variable strength (see Table~\ref{tab:CO-obs}, 
\cite{1981ApJ...248..638L}). Comparison with the brightness curve revealed that the CO emission 
was strongest when the star was around maximum brightness \cite{2006ApJ...651.1130G}. 
Furthermore, the CO emission appeared at the stellar systemic velocity, whereas the absorption 
was either blue- or red-shifted. This behavior is very similar to what has been observed for the 
YHG star $\rho$~Cas \cite{2006ApJ...651.1130G}. Since about 1988, the CO features have 
disappeared from the NIR spectra of V509~Cas, and our spectrum also shows no traces of 
molecular emission. The absorption lines of the hydrogen Pfund series seen in our otherwise 
featureless $K$-band spectrum are in agreement with a stellar temperature of about 8000\,K.

\paragraph{\bf HD~179821 (= RAFGL~2343,  	IRAS 19114+0002, V1427 Aql)} 

The star is embedded within a detached, almost spherical shell of cold (120--140\,K) 
dust \cite{1999ApJ...525L.113J} responsible for an intense far-IR excess 
emission \cite{1989A&A...226..108V}. In the radio regime, CO ($J= 1\longrightarrow  0$) 
emission has been detected, which traces a cold molecular outflow with a velocity of $\sim$33--35\,km\,s$^{-1}$ \cite{1986ApJ...311..345Z}. The object is oxygen-rich, as is inferred 
from its OH maser emission \cite{1989ApJ...344..350L}. 

The evolutionary state of the star is highly debated in the literature due to its uncertain 
distance value. Distance estimates range from about 1.5\,kpc to about 6\,kpc, which would 
classify the star as either a post-asymptotic giant branch star or a YHG. 
Considering the newest parallax measurements of $0.1893\pm 0.0206$ provided by the \textsl{GAIA} Early 
Data Release 3 \cite{2020yCat.1350....0G}, a distance closer to the upper value seems to be 
more likely. Such a high value is also in agreement with the star's kinematic distance 
derived from its large heliocentric systemic velocity of 84--88\,km\,s$^{-1}$
\cite{1986ApJ...311..345Z, 1989ApJ...344..350L, 2008BaltA..17...87K}.

The spectral classification of HD~179821 has been rather controversial as well in the past 
decades  {ranging from} F3--5 
\cite{1996MNRAS.282.1171Z, 1999AJ....117.1834R, 2008BaltA..17...87K}, {over} G5 
\cite{1989ApJ...346..265H}  {to} K4 \cite{1986ApJ...307..711O}. The value for 
$\log g$, derived from high-resolution spectroscopy, ranges around $0.5\pm 0.5$. Such a low $\log g$ 
value assigns the star a luminosity class I. 
%\cite{1996MNRAS.282.1171Z, 1999AJ....117.1834R, 2008BaltA..17...87K, 2018AstL...44..457I}.

%\cite{1986ApJ...307..711O, 2018AstL...44..457I}. 
Extinction values obtained for HD~179821 range from $A_{V} = 2.0$ \cite{1989ApJ...346..265H} over 
$3.1$ \cite{2001AstL...27..156A} to $\sim$4 \cite{1999AJ....117.1834R}, and it has been suggested 
that the total extinction might be variable due to possible changes in the circumstellar contribution 
from the dust shell \cite{2009AstL...35..764A}. 

% \cite{2019A&A...631A..48V}.
 
In a recent work, data from long-term photometric and spectroscopic monitoring have been 
presented \cite{2018AstL...44..457I}. The colors imply that the star first became bluer  
between 1990 and 1995 and has displayed a systematic reddening since 2002. The fastest change in 
color took place between 2013 and 2017, with a simultaneous brightening of the star. The spectra 
confirm this trend and record a more or less stable temperature of $T_{\rm eff} = 6800\pm 
100$\,K between 1994 and 2008, in agreement with previous temperature 
determinations \endnote{We note that a higher effective temperature of $\sim 
7350$\,K was proposed in the same period \cite{2016MNRAS.461.4071S}.} {\cite{1996MNRAS.282.1171Z, 1999AJ....117.1834R, 2008BaltA..17...87K, 2018AstL...44..457I}}. Since then, it decreased 
and reached a value of about 5900\,K in 2017 \cite{2018AstL...44..457I}. The reddening and 
change in temperature indicate the onset of a possible new red loop evolution of HD~179821, i.e., 
an excursion to the cool edge of the HR diagram, related to a significant increase in the 
stellar radius and an increase in mass loss.

In the NIR, HD~179821 displayed CO-band emission, during observations taken between 1989 and 
1991 (see Table~\ref{tab:CO-obs}), whereas no indication for CO band features (neither emission 
nor absorption) was seen between 1992 and 2000. The presence and disappearance of CO band 
emission might indicate a prior phase of higher mass loss or some mass ejection episode 
followed by the subsequent expansion and dilution of the released circumstellar material. 
Such a scenario might be supported by the redder color of the star around \mbox{1990 
\cite{2018AstL...44..457I}.} The absence of CO bands in the spectra between 1992 and 2000 
supports the classification of the star as early to mid F-type, because stars in this 
temperature range are too hot for the formation of molecules in their atmospheres.

\begin{figure}[H]
%\centering
\includegraphics[width=\textwidth]{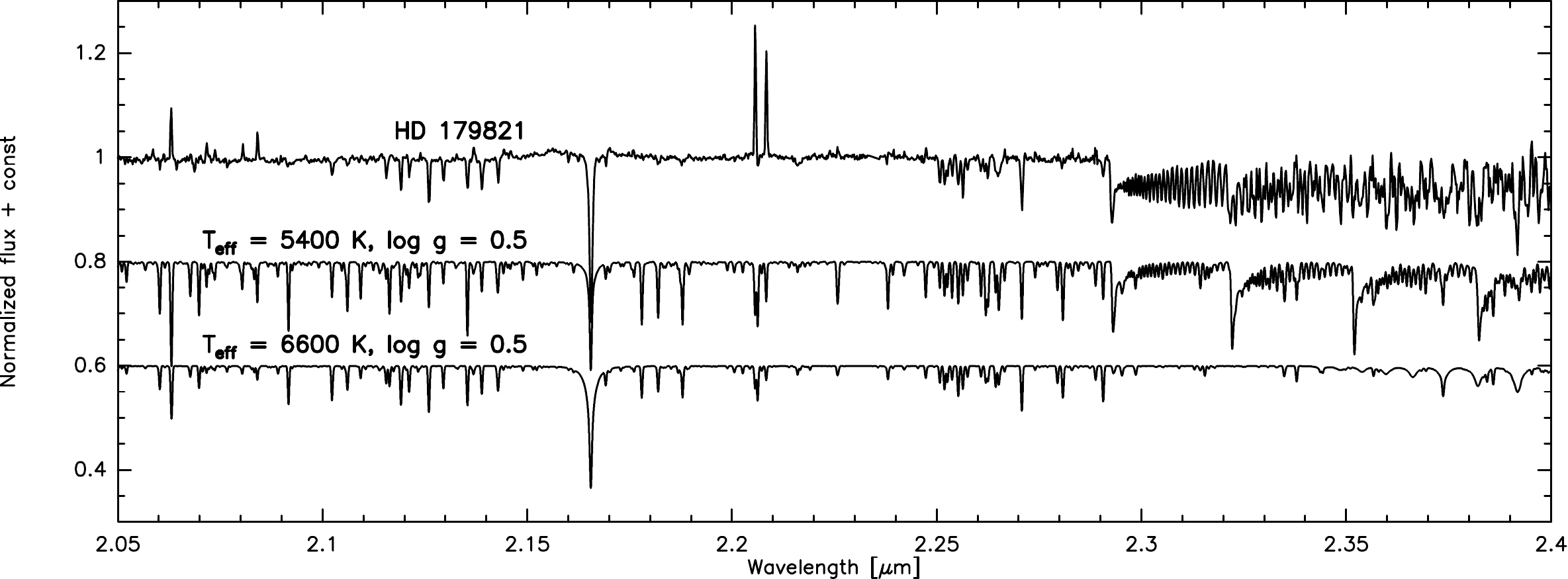}
\caption{Comparison of the $K$-band spectrum of HD~179821 (\textbf{top})
with synthetic spectra for effective temperatures of 5400\,K (\textbf{middle}) and 6600\,K (\textbf{bottom}).
For illustration purposes, the synthetic model spectra are included in this figure and shifted 
down along the flux axis for better visualization.
 \label{fig:HD}}
\end{figure}

In contrast to all previous NIR observations, our data from 2021 clearly display CO band 
absorption. One possible explanation could be that the trend of cooling has continued since 
2017. When comparing our observed spectra with synthetic spectra, we found that the intensity 
of the first band head of CO can be achieved for a stellar effective temperature of $\sim$5400\,K (see Figure~\ref{fig:HD}), although the entire CO band structure signals a 
considerably cooler temperature for the molecular gas due to the only weakly pronounced higher 
band heads. Hotter stars display less intense and cooler stars more intense CO bands. However, 
a star with a temperature of $5400$\,K should show significantly stronger absorption in all 
other atomic photospheric lines, which is not the case. Instead, the intensity and specific 
line ratios in the $K$-band spectrum are more in line with an effective temperature of about 
6600\,K. But such a hot stellar photosphere contains no CO molecular absorption features 
(Figure~\ref{fig:HD}). Based on this discrepancy, we believe that our $K$-band spectrum is 
composite. It shows a hotter stellar photosphere along with CO absorption formed in a 
presumably cooler gas shell or outflow. Support for such a scenario is provided by the fact 
that the CO absorption bands display a blue-shift of about $-43\pm 1$\,km\,s$^{-1}$ with 
respect to all other photospheric atomic lines and the circumstellar emission lines, such as 
the Na\,{\sc i} doublet. If interpreted as outflow velocity, then this value is comparable to 
the outflow seen in IRC~+10420 \cite{1994A&A...281L..33O, 1996MNRAS.280.1062O} but is slightly 
higher than the isotropic expanding cold molecular gas and dust shell around HD~179821 ($\sim$35\,km\,s$^{-1}$, \cite{1986ApJ...311..345Z, 2007A&A...465..457C}). We exclude a cool binary 
component as an explanation for the velocity-shifted CO absorption bands because, in this case, 
the cool companion would imprint (besides CO) significantly stronger blue-shifted absorption 
lines onto the $K$-band spectrum, which are not observed. If the proposed scenario of a new 
outflowing shell or gas layer is correct, possibly initiated during the reddening of the star 
and the increase in stellar radius recorded in the years 2014--2017 \cite{2018AstL...44..457I}, 
or a possible outburst event that might have followed this reddening as in the case of V509~Cas
(as we mentioned before, \mbox{see \cite{2012A&A...546A.105N, 1975ApJ...202..743L}),}
then we may speculate that with further expansion of this matter, future $K$-band observations 
will display CO bands in emission before the material dilutes and the CO features might disappear 
again.

Besides CO, our spectrum of HD~179821 also shows Br\,$\gamma$ in absorption and intense 
emission of the Na{\sc i} doublet. The line of Br\,$\gamma$ was in absorption in previous 
observations taken in 1989 \cite{1995A&A...299...69O} and 1990 \cite{1994ApJ...420..783H}. The 
latter spectrum also displays intense emission of the (blended) Na{\sc i} doublet, as well as 
the spectrum taken in 2000 \cite{2006ApJ...651.1130G}. The remaining previous spectra either do 
not cover the spectral region of Br\,$\gamma$ and/or the Na{\sc i} doublet, or these lines 
have not been mentioned by the corresponding authors.

%\unskip

Na\,{\sc i} emission is a clear indicator for circumstellar material. It has been reported from
the NIR spectra of numerous evolved massive stars: (i) the YHGs $\rho$~Cas, V509~Cas, 
Hen~3-1379 (the Fried Egg Nebula), IRC~+10420 \cite{1981ApJ...248..638L, 2022MNRAS.515.2766K}, 
and the IRC~+10420 analog IRAS~18357-0604 \cite{2014A&A...561A..15C}, (ii) most of the B[e]SGs 
\cite{1988ApJ...334..639M, 2013A&A...558A..17O, 2021BAAA...62..104A}, and (iii) also many 
luminous blue variables \cite{2013A&A...558A..17O, 2017A&A...601A..76K}. The equivalent widths 
of the intense Na\,{\sc i} lines from the YHG IRC~+10420 \cite{2013A&A...551A..69O} are about 
three times higher than for HD~179821, for which we measured values of $-1.0670\pm 0.015$\,\AA \
and $-0.9231\pm 0.018$\,\AA. Recent spatially resolved observations revealed that the Na\,{\sc i} 
emission in IRC~+10420 and Hen~3-1379 is confined within a compact spherical envelope around 
the star \cite{2022MNRAS.515.2766K, 2020A&A...635A.183K}. The emission lines in our spectrum 
are symmetric, and their wavelengths coincide with the systemic velocity of the star. While 
their formation region can be a compact spherical shell as well, they cannot be related to the 
possible new blue-shifted outflow traced by the CO band absorption.

%%%%%%%%%%%%%%%%%%%%%%%%%%%%%%%%%%%%%%%%%%
\section{Conclusions}

We present new medium- and high-resolution $K$-band spectra for two B[e]SGs and four YHGs. The spectra 
of both B[e]SGs show rotationally broadened CO band emission, from which we could derive, for the
first time, the projected rotation velocity of the CO gas for both stars. On the other hand, our 
model parameters for the CO temperature and column density are very similar to those reported in 
previous studies based on spectra with significantly lower resolution \cite{2010MNRAS.408L...6L, 
2013A&A...558A..17O}. The similarities of the detected CO band features over more than 30 years
suggest that the CO emitting gas rings around these two B[e]SGs are stable structures, neatly fitting 
to the findings of most of the B[e]SGs.

With respect to the YHGs, we detect CO band emission from only one star, the highly reddened cluster 
member [FMR2006]~15, which previously showed time-variable CO features (see Table~\ref{tab:CO-obs}), 
and which has an effective temperature that is clearly too high to form molecules within its 
atmosphere. Consequently, the CO emission must be of circumstellar origin. A second object, HD 179821, 
shows CO bands in absorption, while it had CO emission during 1989--1991 but lacked any CO features in 
its spectrum since 1992 (Table~\ref{tab:CO-obs}). The latter is consistent with the star's high 
effective temperature, which prevents the formation of molecules. For both YHGs, our detected CO 
features are clearly blue-shifted with respect to the photospheric absorption lines, suggesting that 
both stars most likely had recent mass ejection events and the CO  {emission/absorption} 
forms within the expelled matter. For [FMR2006]~15, we propose that the blue-shifted emission
arises in a possible pole-on seen bipolar outflow as in the case of the YHG star IRC~+10420 
\cite{2013A&A...551A..69O}, but nothing can be said yet about the geometry of the outflow seen from 
HD~179821 because the (highly inflated) star itself might still block large portions of the possibly 
receding parts of \linebreak the ejecta.

\vspace{6pt} 

%%%%%%%%%%%%%%%%%%%%%%%%%%%%%%%%%%%%%%%%%%
%% optional
%\supplementary{The following supporting information can be downloaded at:  \linksupplementary{s1}, Figure S1: title; Table S1: title; Video S1: title.}

%%%%%%%%%%%%%%%%%%%%%%%%%%%%%%%%%%%%%%%%%%
\authorcontributions{Conceptualization, M.K. (Michaela Kraus), M.K. (Michalis Kourniotis) and D.H.N.; methodology, M.K. (Michaela Kraus), M.K. (Michalis Kourniotis), %Kraus: adjusted
M.L.A. and A.F.T.; formal analysis, M.K. (Michaela Kraus) and M.K. (Michalis Kourniotis); investigation, M.K. (Michaela Kraus), M.K. (Michalis Kourniotis) and D.H.N.; resources, 
M.K. (Michaela Kraus), M.K. (Michalis Kourniotis), M.L.A. and A.F.T.; writing---original draft preparation, review and editing, M.K. (Michaela Kraus), %Kraus: adjusted
M.K. (Michalis Kourniotis), M.L.A., A.F.T. and D.H.N.; visualization, M.K. (Michaela Kraus) and M.K. (Michalis Kourniotis); funding acquisition, M.K. (Michaela Kraus), M.L.A. 
and A.F.T. All authors have read and agreed to the published version of the manuscript.}

%%%%%%%%%%%%%%%%%%%%%%%%%%%%%%%%%%%%%%%%%%
\funding{This research was funded by the Czech Science foundation (GA \v{C}R, grant number 20-00150S),
by CONICET (PIP 1337), by the Universidad Nacional de La Plata (Programa de Incentivos 11/G160), 
Argentina, and by the European Union's Framework Programme for Research and Innovation Horizon 2020 
(2014-2020) under the Marie Sk\l{}odowska-Curie Grant Agreement No. 823734. The Astronomical Institute 
of the Czech Academy of Sciences is supported by the project RVO:67985815.}

\institutionalreview{Not applicable.}

\informedconsent{Not applicable.}

\dataavailability{The data underlying this article will be shared on reasonable request
to the corresponding author.
%{\it In this section, please provide details regarding where data supporting reported results can be found, including links to publicly archived datasets analyzed or generated during the study. Please refer to suggested Data Availability Statements in section ``MDPI Research Data Policies'' at \url{https://www.mdpi.com/ethics}. If the study did not report any data, you might add ``Not applicable'' here.}
} 

%%%%%%%%%%%%%%%%%%%%%%%%%%%%%%%%%%%%%%%%%%
\acknowledgments{{We thank the anonymous referees for their valuable comments and suggestions.} 
This research made use of the NASA Astrophysics Data System (ADS) and of the SIMBAD database, operated 
at CDS, Strasbourg, France. This paper is based on observations obtained with the Phoenix infrared
spectrograph, developed and operated by the National Optical Astronomy Observatory and based on 
observations obtained at the international Gemini Observatory, a program of NSF’s NOIRLab, which is 
managed by the Association of Universities for Research in Astronomy (AURA) under a cooperative 
agreement with the National Science Foundation on behalf of the Gemini Observatory partnership: the 
National Science Foundation (United States), National Research Council (Canada), Agencia Nacional de 
Investigaci\'{o}n y Desarrollo (Chile), Ministerio de Ciencia, 
Tecnolog\'{i}a e Innovaci\'{o}n (Argentina), Minist\'{e}rio da Ci\^{e}ncia, 
Tecnologia Inova\c{c}\~{o}es e Comunica\c{c}\~{o}es (Brazil), and Korea Astronomy and Space 
Science Institute (Republic of Korea) under program IDs GS-2004B-Q-54, GS-2017B-Q-32, 
GN-2019A-Q-204, GN-2019B-Q-418, and GN-2021A-Q-315.
}

%%%%%%%%%%%%%%%%%%%%%%%%%%%%%%%%%%%%%%%%%%
\conflictsofinterest{The authors declare no conflict of interest.} 
%The funders had no role in the design of the study; in the collection, analyses, or interpretation 
%of data; in the writing of the manuscript, or in the decision to publish the results.} 

%%%%%%%%%%%%%%%%%%%%%%%%%%%%%%%%%%%%%%%%%%
\newpage
\abbreviations{Abbreviations}{
The following abbreviations are used in this manuscript:\\

\noindent
\begin{tabular}{@{}ll}
B[e]SG & B[e] supergiant\\
NIR & Near infrared\\
LMC & Large Magellanic Cloud\\
SMC & Small Magellanic Cloud\\
YHG & Yellow hypergiant
\end{tabular}
}

%%%%%%%%%%%%%%%%%%%%%%%%%%%%%%%%%%%%%%%%%%
\begin{adjustwidth}{-\extralength}{0cm}

\printendnotes[custom]

\reftitle{References}

\PublishersNote{}
\end{adjustwidth}
\end{document}